\documentclass{elsart1p}

\usepackage{epsfig}

\usepackage{amssymb}

\begin{document}

\begin{frontmatter}



\title{Mass and chemical asymmetry in QCD matter}


\author[a]{L. F. Palhares\corauthref{*}},
\corauth[*]{Corresponding author.}
\ead{leticia@if.ufrj.br}
\author[a]{E. S. Fraga}, and
\author[b]{C. Villavicencio}

\address[a]{Instituto de F\'\i sica, Universidade Federal do Rio de Janeiro, \\
C.P. 68528, Rio de Janeiro, RJ 21941-972, Brazil}
\address[b]{Departamento de F\'\i sica, Comisi\'on Nacional de Energ\'\i a At\'omica, (1429) Buenos Aires, Argentina}

\begin{abstract}

We consider two-flavor asymmetric QCD combined with a
low-energy effective model inspired by chiral perturbation theory
and lattice data to investigate the effects of masses, isospin and baryon number
on the pressure and the deconfinement phase transition. Remarkable agreement
with lattice results is found for the critical temperature behavior. Further
analyses of the cold, dense case and the influence of quark mass asymmetry
are also presented.
\end{abstract}

\end{frontmatter}


The phase diagram of strong interactions embraces at the same time
several physical phenomena and theoretical open questions and has been
intensively studied during the last years.
This investigation involves not only the full usage of in-medium techniques
but also the understanding of the non-perturbative regime of QCD,
lattice simulations being the main tool to probe the full theory
\cite{Laermann:2003cv}.

Despite the fact that Lattice QCD is increasingly providing results with realistic
(small) quark masses and probing a larger domain at finite baryon 
chemical potential, it is still considerably restricted by 
current lattice sizes and the Sign Problem \cite{Cheng:2007jq,Philipsen:2008gf}. 
QCD at finite isospin density and vanishing baryon chemical potential
provides a useful framework to test alternative techniques to simulate finite
density on the lattice, since it has a positive fermionic determinant and therefore
no Sign Problem. In addition, it is also part of the physical phase diagram 
for strong interactions, and exhibits a very rich phenomenology \cite{Son:2000xc}.
Theoretical predictions with 
finite quark masses, isospin and baryon densities are then crucial for 
both technical and phenomenological reasons, complementing in a natural way
the full QCD thermodynamics probed by lattice simulations.

We investigate the effects of finite quark masses, isospin, and baryon 
number on the equation of state of QCD matter and on the 
deconfining phase transition within an effective model inspired by lattice 
results and chiral perturbation theory ($\chi$PT). Results for the
critical temperature \cite{Fraga:2008be} are compared to lattice data showing remarkable
agreement in contrast with previous model predictions. The influence of the 
quark-mass asymmetry on the deconfinement transition at finite temperature 
is also analyzed. Furthermore, we present results for the mass and isospin 
dependence of the critical baryon chemical potential \cite{future}.
While the high temperature regime is of interest to the phenomenology of 
high-energy heavy ion collisions, the findings for cold, dense matter can
be relevant for compact stars and for the 
phenomenon of color superconductivity \cite{Glendenning:1997wn}.


To study the influence of masses and chemical potentials on the deconfining 
phase transition, we constructed a simple effective model with explicit 
breaking of the chiral and isospin symmetries, including the relevant 
parameters (quark masses $m_u$ and $m_d$, and quark chemical potentials 
$\mu_u = (\mu_B+\mu_I)/2$ 
and $\mu_d=(\mu_B-\mu_I)/2$) 
in the high and low energy regimes in a consistent manner \cite{Fraga:2008be}.
This feature induces an interdependent implementation of variations 
of quark masses and chemical potentials in both sectors that percolates
in a positive fashion to the critical parameters.


The low energy regime is described by a $\chi$PT-inspired effective model of 
dressed pions and nucleons. The quasi-pions satisfy effective dispersion relations
which incorporate temperature and isospin chemical potential corrections, as well 
as the dependence on the quark masses, $m_{u/d}=m\mp \delta m$. They were calculated for both 
phases $\mu_I<m_{\pi}$ and $\mu_I> m_{\pi}$ (in which the negatively charged 
pions Bose condense) in Refs. \cite{Loewe:2002tw}.
The nucleon effective masses $M_{p/n}= M\mp \delta M$ are taken to be 
the leading-order result in zero-temperature Baryon-$\chi$PT as functions of 
the quark average mass, $m$, and mass difference, $\delta m$ \cite{Procura&Beane}.


For the high energy sector, we adopt 2-flavor QCD with massive quarks 
and explicit isospin symmetry breaking ($m_d, m_u \ne 0$, $m_u \ne m_d$) 
and complement phenomenologically the perturbative result with non-perturbative 
corrections through the fuzzy bag model \cite{Pisarski:2006hz}. Besides the usual MIT-type bag 
constant, the total QCD pressure in this model has also a non-perturbative thermal contribution
$\sim T^2$ to account for the unusually flat behavior of the trace anomaly
normalized to $T^2$ observed in lattice results above the critical temperature.
Here we adopt a simple extension of the model introduced in \cite{Pisarski:2006hz}
which includes finite masses and chemical potentials within the perturbative contribution.

All the coefficients of our effective model are either fixed to reproduce observed properties 
of the QCD vacuum or extracted from lattice simulations. For more details on 
the model, explicit expressions and parameter fixing through lattice data 
and vaccuum QCD observables, the reader is referred to \cite{Fraga:2008be,future}.


We computed the massive free gas contribution of the pQCD pressure in the 
fuzzy bag model at finite temperature, isospin and baryon number and the 
free gas pressure of quasi-pions and nucleons in the low energy regime \cite{Fraga:2008be}.
The critical temperature and chemical potential for the deconfining phase 
transition are then extracted by maximizing the total pressure. It should 
be noticed that the results for very large pion masses are in principle outside 
the domain of validity of our approximations, since the low energy regime 
of the model is based on leading-order $\chi$PT. It is still interesting however 
to determine the qualitative tendency.

\begin{figure}[htb]
\begin{minipage}[t]{64mm}
\includegraphics[width=6.5cm]{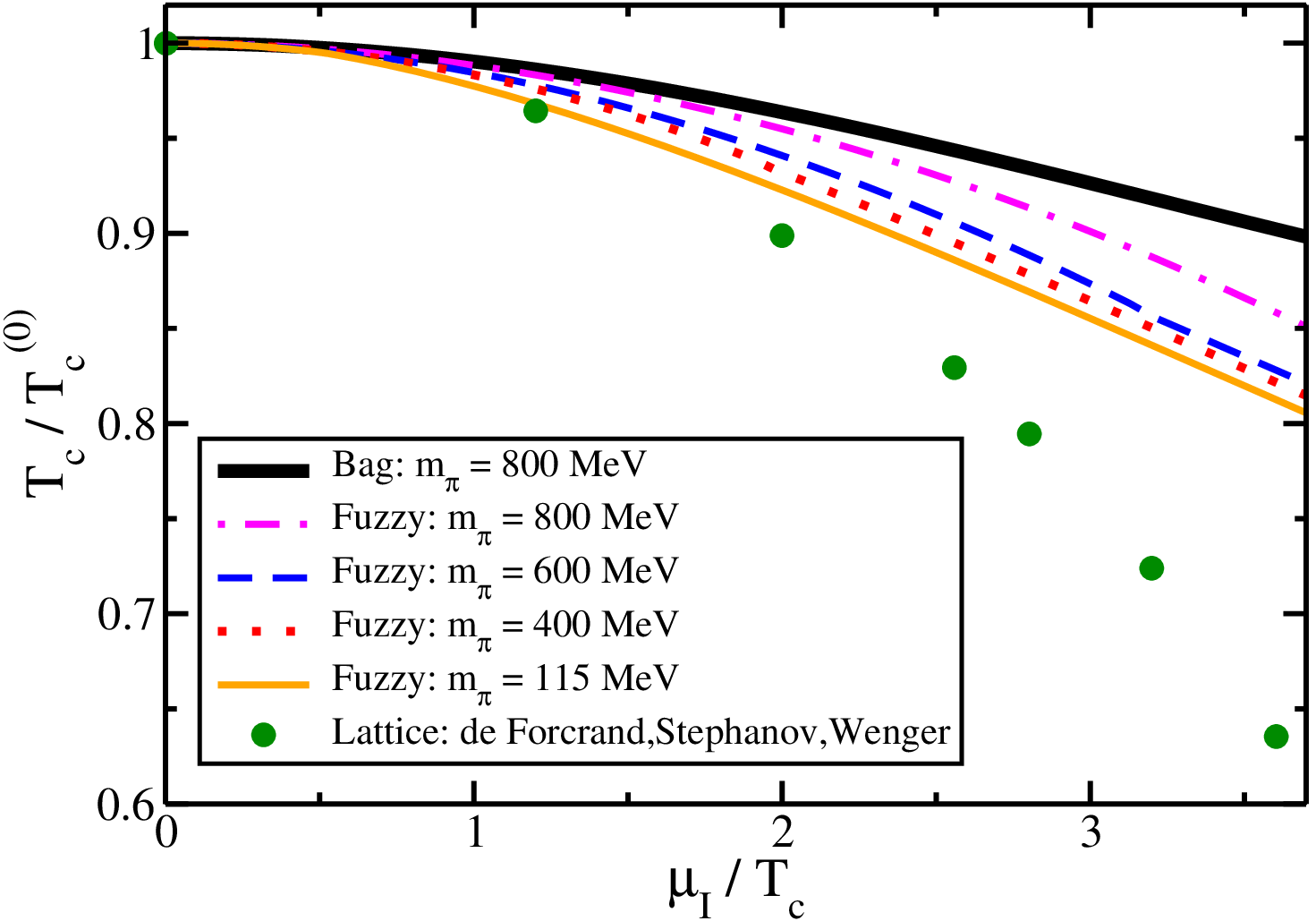}
\caption{Critical temperature as a function of $\mu_I$ for different 
values of $m_{\pi}$. Lattice data from \cite{deForcrand:2007uz}.}
\label{Tc-muI}
\end{minipage}
\hspace{.1cm}
\begin{minipage}[t]{70mm}
\includegraphics[width=6.3cm]{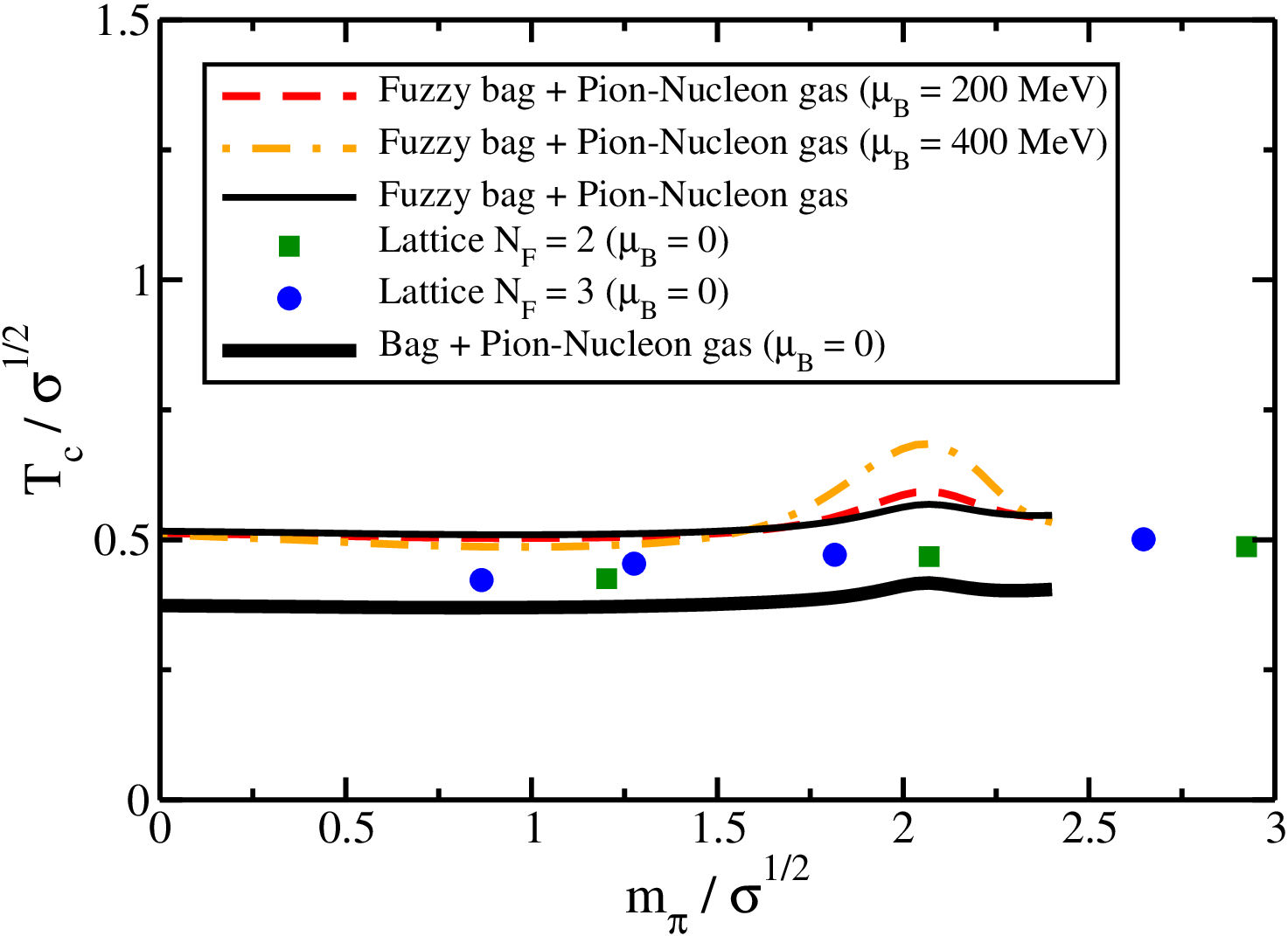}
\caption{Critical temperature as a function of $m_\pi$ normalized by 
the string tension $\sqrt{\sigma}=425~$MeV, for different $\mu_B$
($\mu_I=0$). Lattice data from \cite{Karsch:2000kv}.}
\label{Tc-mpi}

\end{minipage}
\end{figure}

In Fig. \ref{Tc-muI}, the critical temperature is plotted as a function of the isospin 
chemical potential. The results are in very good agreement with lattice 
computations \cite{deForcrand:2007uz}, even though the curves closer to the lattice points 
correspond to smaller vacuum pion masses, which is not the situation  
simulated on the lattice. In comparison with previous results, our model appears 
to agree with lattice data within a considerably larger interval of isospin chemical 
potential \cite{Fraga:2008be}. 

\vspace{0.65cm}

The pion mass dependence, or equivalently the quark average mass dependence,
of the critical temperature is displayed in Fig. \ref{Tc-mpi}. The 
approximate mass independence observed in the lattice data \cite{Karsch:2000kv} is well 
reproduced here, while previous treatments tended to generate a rather different 
behavior (cf. Fig. 1 in Ref. \cite{Fraga:2008be}). The effect of a finite $\mu_B$ is
completely imperceptible for $\mu_B\le 100~$MeV. We also computed the critical 
parameters by considering the usual bag model in the high energy regime,
finding values systematically lower, but with the same qualitative behavior
as with the fuzzy bag. On the other hand, the behavior of $T_c$ with $\mu_I$ for 
the effective model with the usual bag pressure
does not reproduce lattice data as well as the fuzzy one: even for $m_{\pi}=25~$MeV, the bag model
curve is still between the $m_{\pi}=400~$MeV and $m_{\pi}=600~$MeV fuzzy results.

\vspace{0.4cm}
\begin{figure}[htb]
\begin{minipage}[t]{67.5mm}
\includegraphics[width=6.75cm]{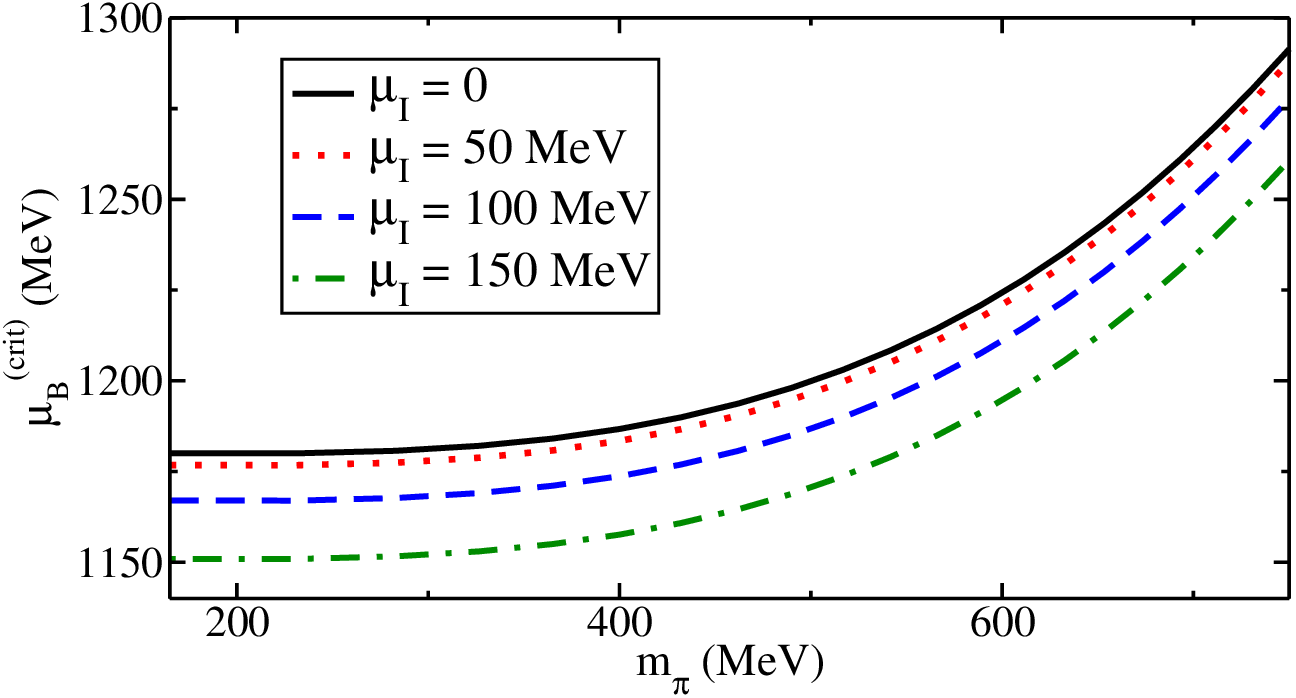}
\caption{Critical baryon chemical potential as a function of $m_\pi$, for different values of the isospin chemical potential.}
\label{muBc-mpi}
\end{minipage}
\hspace{.1cm}
\begin{minipage}[t]{66mm}
\includegraphics[width=6.6cm]{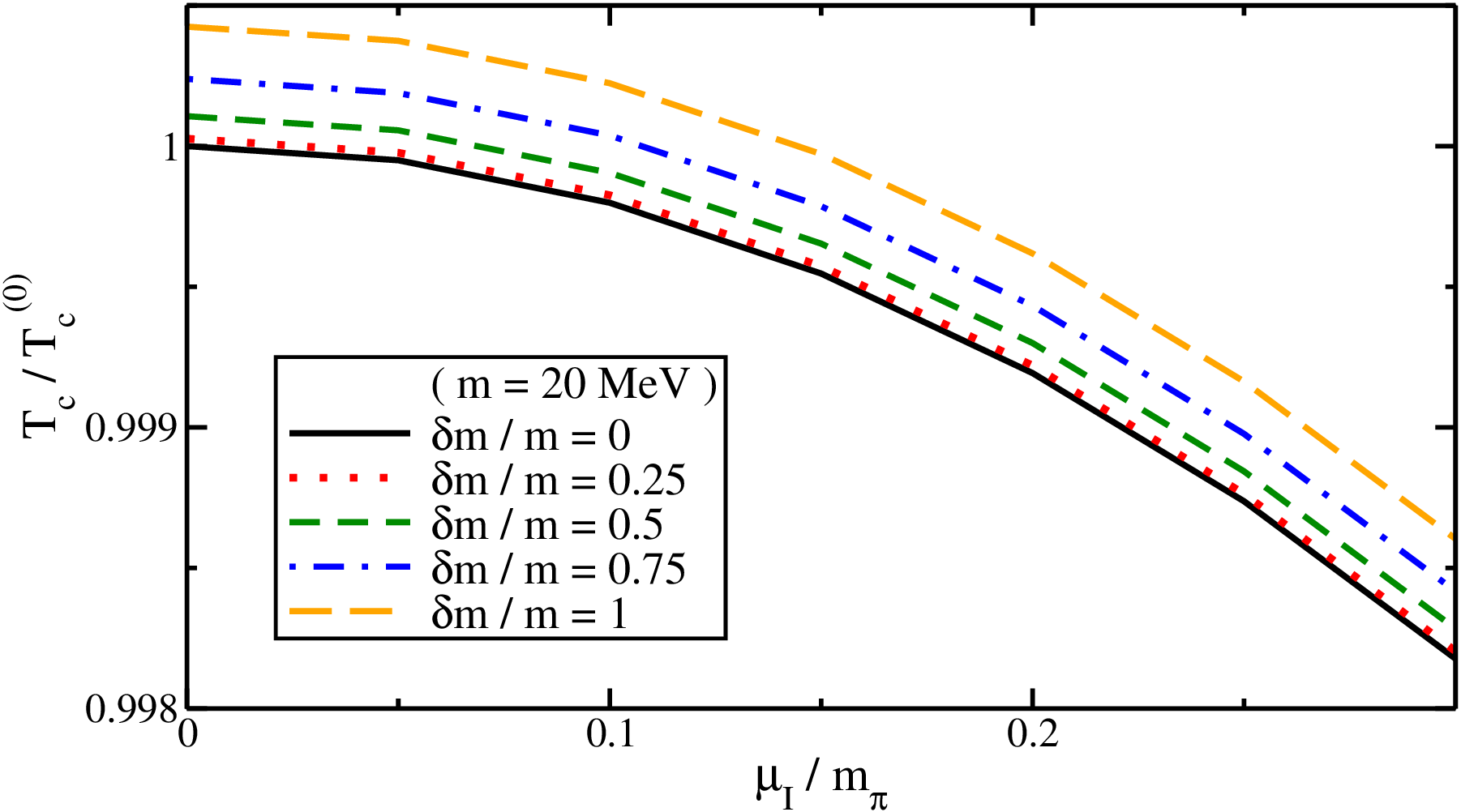}
\caption{Critical temperature as a function of $\mu_I$ for different values of the quark mass difference $\delta m$.}
\label{Tc-deltam}
\end{minipage}
\end{figure}

Fig. \ref{muBc-mpi} displays the cold, dense case: the value of baryon chemical 
potential beyond which matter is deconfined increases with the vacuum 
pion mass. The outcome of a finite, small isospin density is to decrease 
the critical baryon chemical potential. This property could in principle 
play an important role in astrophysics.
It is interesting to notice that the critical baryon chemical potential seems to be
significantly more sensitive to mass variations in comparison with the finite temperature
deconfinement transition. The relevance of mass effects in the thermodynamics of cold,
dense matter has already been shown in different related contexts, such as perturbative
QCD \cite{Fraga:2004gz,Laine:2006cp} and perturbative Yukawa theory \cite{Palhares:2008yq}.

The effect of considering a finite quark mass difference is presented in 
Fig. \ref{Tc-deltam}. It tends to increase the critical temperature, though by a 
quantitatively small amount, as expected.


We constructed a simple effective model that includes all the relevant ingredients
with a clear and consistent connection between the high and low energy 
parameters, allowing for a systematic investigation of the influence of 
finite quark masses and isospin symmetry breaking on the QCD phase diagram at 
finite temperature and density. 
Despite the simplicity of the framework, a free gas equation
of state yielded the critical deconfinement temperature as a function of
the pion mass and the isospin chemical potential in surprisingly good 
agreement with different sets of lattice data. This success indicates that the
model captures the essential features associated with nonzero masses and chemical
potentials. For the mass dependence, the heavy nucleon content in the low energy
sector is probably crucial, since the previous approaches that failed to render the
approximate mass invariance of the critical temperature consisted essentially of chiral fields.
On the other hand, the behavior of $T_c(\mu_I)$ suggests that the quarks and 
the quasi-particles at low energy represent the adequate degrees of freedom 
carrying isospin charge in each phase. Furthermore, this model can 
supply information about the cold and dense regime of QCD matter, which is
not yet fully covered by lattice simulations.

\vspace{0.1cm}


\noindent{\bf Acknowledgements:} 
This work was partially supported by ANPCyT, CAPES, CNPq, FAPERJ,
and FUJB/UFRJ.



\end{document}